\documentclass{article}
\usepackage{amsfonts,amsmath,amssymb}
\date{}
\def\integer{\mathbb Z}
\def\real{\mathbb R}
\title{Exactly solvable periodic Darboux $q$-chains
\thanks{This work was partially supported by Russian Foundation 
for Fundamental Research (grants \char"C2~02-01-00659, 00-15-96011)}
}
\author{Ivan A.\,Dynnikov and Sergey V.\,Smirnov}

\begin{document}
\maketitle
\noindent\smash{\raisebox{4cm}{Russian Math.\ Surveys 57:4 (2002), to appear}}

Let $L_1,L_2,\dots$ be selfadjoint differential operators acting on 
$\mathbb R$. They form {\it a Darboux chain} if they satisfy the relation
\begin{equation}\label{darbouxchain}
L_j=A_jA_j^+-\alpha_j=A_{j-1}^+A_{j-1},
\end{equation}
where $A_j=-d/dx+f_j(x)$ are first order differential operators. A Darboux 
chain is called \emph{periodic} if $L_{j+r}=L_j$ for some $r$ and for all 
$j=1,2,\dots$. Number $r$ is called {\it the period} of a Darboux chain. 
The operator $L+\frac\alpha2$ appears to be the harmonic oscillator in 
the particular case $r=1$.

Periodic Darboux chains lead to integrable systems of differential
equations for functions $f_j$, which are examined
in~\cite{VS}. The cases $\alpha=0$ and
$\alpha\ne0$, where $\alpha=\sum_{j=1}^r\alpha_j$, are cardinally different. 
The operators of a periodic Darboux chain are finite-gap if $\alpha=0$.
If $\alpha\ne0$, then the equations for $f_j$ lead to the Painlev\'e equations 
or their higher analogues; relations~(\ref{darbouxchain}) define the 
discrete spectrum of the chain operators $L_j$: the spectrum of each of these 
operators consists of $r$ arithmetic sequences (see~\cite{VS}).

Consider the following ``$q$-analogue'' of a Darboux chain:
\begin{equation}\label{qchain}
L_j=A_jA_j^+-\alpha_j=qA_{j-1}^+A_{j-1},
\end{equation}
where $A_j=a_j+b_jT$ are difference operators on one-dimensional lattice 
$\integer$ and $T$ is the shift operator: $(A_jf)(n)=a_j(n)f(n)+b_j(n)f(n+1)$,
$a_j(n),b_j(n)\in\real\backslash\{0\}$. Without loss of generality we may
assume that $b_j(n)>0$ for all $j,n$. One of the
important features of difference
operators in comparison to differential ones is the possibility to define a 
periodic chain in various ways. We say that a 
chain~(\ref{qchain}) is {\it periodic with period $r$ and shift $s$} if 
the relation
\begin{equation}\label{rs}
L_{j+r}=T^{-s}L_jT^s
\end{equation}
holds for all $j\geqslant 1$. In the literature the case $s=0$ has mostly 
been considered. However the possibility of factorization in 
``the reverse order'' was mentioned in~\cite{NT,ND}: 
operators $A_j=a_j+b_jT$ can be replaced by operators of the form 
$A_j=a_j T^++b_j$; such factorization with $s=0$
is, in fact, equivalent to the
factorization ``in the right order'' for $s=r$
(the operator $L_j$ is replaced
by $T^{-j}L_jT^j$). Apparently, the general formulation
with an arbitrary $s$
has not been considered before, while
the following special cases have been studied in the literature.

\smallskip\noindent1. $\alpha=0$, $q=1$, $r$ is arbitrary~\cite{NT}.
Operators $L_j$ are finite-gap.

\smallskip\noindent2. $\alpha>0$, $q=1$, $r=1$ (difference analogue of the 
harmonic oscillator)~\cite{NT}. In this case, symmetric operators $L_j$ 
acting on the space of functions on the lattice $\integer$ do not exist. 
Nevertheless, there exists a solution on the ``half-line'' $\integer_{>0}$.
The spectrum of the operator $L+\frac\alpha2$ is exactly the same as that 
of the harmonic oscillator: $\lambda_k=\alpha(k+\frac12)$; the 
eigenfunctions are expressed in terms of the Charlier polynomials, and, 
therefore, form a complete family in $\mathcal L_2(\integer_{>0})$.

\smallskip\noindent3. $r=1$, $\alpha>0$, $0<q<1$ (or $\alpha<0$,
$1<q$)~\cite{NT,SVZ} ($q$-oscillator). The spectrum of the operator $L$ lies 
in the interval $[0,\frac{q\alpha}{1-q})$ (or in 
$(0,\frac{q\alpha}{q-1}$ if $q>1$); it forms a ``$q$-arithmetic sequence''.
It is mentioned in~\cite{NT} that, in this case, the operator $L$ is 
unbounded, and cojectured that $L$ has continuous spectrum in the interval 
$(\frac{q\alpha}{1-q},\infty)$.

\smallskip\noindent4. Another version of the $q$-oscillator is considered 
in~\cite{AFW}. It is presented by a difference operator on the 
whole ``line'' $\integer$. In our settings,
this version of the $q$-oscillator can be
interpreted as the case $s=1$, $r=2$, $\alpha_1=\alpha_2$, ($\alpha$ and $q$ 
are the same as in 3). A particular 
solution that is symmetric about the origin was found in~\cite{AFW}.
This case is specific because here the operator $L$ is bounded and has no 
continuous spectrum.

\smallskip
We claim here that the same property holds for a $q$-chain of an arbitrary 
even period $r$ with the shift $s=r/2$: the chain operators are bounded and 
have no continuous spectrum. We also provide explicitly the general solution 
of the problem in the case $s=1$, $r=2$.

\medskip
\noindent{\bf Theorem.}
{\it
Suppose $r$ is even, $\alpha_1,\dots,\alpha_r$ are positive, $q$ satisfies 
the inequality $0<q<1$, and we have $s=r/2$. Then the 
system~(\ref{qchain}),(\ref{rs}) 
has an $r$-parametric family of solutions. The operator $L_j$ is bounded for 
each $j$; its spectrum $\{\lambda_{j,0},\lambda_{j,1},\dots\}$ is discrete and 
is contained in the interval $[0,\|L_j\|)$. It can be found by using 
the Darboux scheme:
$$\lambda_{j,0}=0,\quad\lambda_{j+1,k+1}=q(\lambda_{j,k}+\alpha_j),
\quad\lambda_{j+r,k}=\lambda_{j,k}.$$
For each $j$, the eigenfunctions of the operator $L_j$ also can be obtained 
by using the Darboux scheme:
$$A_{j-1}\psi_{j,0}=0,\quad\psi_{j+1,k+1}=A_j^+\psi_{j,k};$$
these eigenfunctions form a complete family in $\mathcal L_2(\integer)$.

A similar assertion holds for $\alpha_1,\dots,\alpha_r<0$, $1<q$ (in this 
case, the point $0$ is not included in the spectrum of $L_j$).}
\medskip

\medskip
Now we present an explicit form of the operators $A_1,A_2$ for $r=2$.

\medskip
\noindent{\bf Proposition.}
{\it
For $r=2=2s$, $\alpha_1,\alpha_2>0$, $0<q<1$, the general solution of 
the problem~(\ref{qchain}),(\ref{rs}) has the form
$a_1(n)=\epsilon\sqrt{\xi_{2n}}$, $b_1(n)=\sqrt{\eta_{2n+1}}$,
$a_2(n)=\epsilon\sqrt{\xi_{2n-1}}$, $b_2(n)=\sqrt{\eta_{2n}}$,
where $\epsilon=\pm1$,
\begin{equation}\label{xy}
\xi_n=\frac12\frac
{c_n-2\kappa\,q^{-n-\varphi-\frac12}+c_{n+1}q^{-2n-2\varphi-1}}
{(1-q^{-2(n+\varphi)})(1-q^{-2(n+\varphi+1)})},\quad
\eta_n=\frac12\frac
{c_{n+1}-2\kappa\,q^{n+\varphi+\frac12}+c_nq^{2n+2\varphi+1}}
{(1-q^{2(n+\varphi)})(1-q^{2(n+\varphi+1)})},
\end{equation}
$$c_n=\frac{\alpha_1+\alpha_2}{1-q}+(-1)^n\frac{\alpha_1-\alpha_2}{1+q},$$
$\varphi\in{\mathbb R}$ is arbitrary, the parameter $\kappa$ satisfies the restrictions
$$c_{[\varphi]}q^{-\theta}+c_{[\varphi]-1}q^\theta
<2\kappa<\min(c_{[\varphi]}q^{\theta+1}+c_{[\varphi]-1}q^{-\theta-1},
c_{[\varphi]}q^{\theta-1}+c_{[\varphi]-1}q^{-\theta+1})\quad
\mbox{if }\varphi\notin\integer,$$ and 
$2\kappa=c_\varphi q^{\frac12}+c_{\varphi-1}q^{-\frac12}$ if 
$\varphi\in\integer$ (in this case, in fractions~(\ref{xy}), one 
has to cancel the factor $(1-q^{2(n+\varphi)})$ for 
$n\equiv\varphi(\mathrm{mod}\,2)$ and the factor $(1-q^{2(n+\varphi+1)})$ for 
$n\equiv\varphi+1(\mathrm{mod}\,2)$). Here, we set 
$\theta=\varphi-[\varphi]-\frac12$ and $[\varphi]$ stands for the integral 
part of $\varphi$.}

\medskip

\noindent{\bf Observation.}
In the case $r=2$, $\alpha_1=\alpha_2$, operators $L_j+\frac{\alpha_j}2$ 
constructed from the above solutions converge to the harmonic 
oscillator as $q\rightarrow1$ in the following sense: take $\varphi=0$, 
$\epsilon=-1$, $q=\exp(-\frac{\alpha_1}4h^2)$,
$x=nh$, $T=\exp(h\frac d{dx})$
and assume that $n$ is real in formulae~(\ref{xy}) and that the operator 
$L_j$ is a difference operator on $\real$. Then for any 
$f\in C^2(\real)$ we have
$$\Bigl(L_{1,2}+\frac{\alpha_1}2\Bigr)f(x)=\Bigl(-\frac{d^2}{dx^2}+
\frac{\alpha_1^2}4x^2\Bigr)f(x)+o(h).$$
If $\alpha_1\ne\alpha_2$, then the operator $L_j$ converges in the same sense to
$$-\frac{d^2}{dx^2}+\frac{(\alpha_j+\alpha_{j+1})^2}{16}x^2
-\frac{\alpha_j}2-\frac{(\alpha_j-\alpha_{j+1})(\alpha_j+3\alpha_{j+1})}
{4(\alpha_j+\alpha_{j+1})^2x^2},$$
where $\alpha_{j+2}=\alpha_j$.

In the cases 2,3 mentioned above there is no link of that kind between the 
discrete and the continuous models. Thus the considered case $s=r/2$ gives,
in a certain sense, a proper discretization of a Darboux chain.

We hope to prove that a $q$-chain converges in the same way to an ordinary 
Darboux chain for an arbitrary even $r$. The numerical experiment 
confirms that a $q$-chain of the period $6$ converges to an ordinary Darboux 
chain of the period $3$ if $\alpha_j=\alpha_{j+3}$.

\end{document}